\documentclass[10pt,twocolumn]{article}
\usepackage[margin=1in]{geometry}
\usepackage[T1]{fontenc}
\usepackage{textcomp}
\usepackage{hyperref}
\usepackage{parskip}
\usepackage{enumitem}
\usepackage{booktabs}
\usepackage{amsmath,amssymb,mathtools}
\usepackage{graphicx}
\usepackage{microtype}
\usepackage{xcolor}
\usepackage{algorithm}
\usepackage{algpseudocode}
\usepackage{subcaption}
\usepackage{multirow}
\usepackage{tikz}
\usetikzlibrary{positioning,shapes,arrows.meta}

\title{\bfseries EigenAI: Deterministic Inference, Verifiable Results}
\author{
David Ribeiro Alves\\ \texttt{david@eigenlabs.org}
\and
Vishnu Patankar\\ \texttt{vishnu@eigenlabs.org}
\and
Matheus Pereira\\ \texttt{matheus@eigenlabs.org}
\and
Jamie Stephens\\ \texttt{jamie@eigenlabs.org}
\and
Nima Vaziri\\ \texttt{nima@eigenlabs.org}
\and
Sreeram Kannan\\ \texttt{sreeram@eigenlabs.org}
}
\date{}

\begin{document}
\maketitle
\vspace*{-12pt}

\begin{abstract}
\noindent
\textbf{EigenAI} is a verifiable AI platform built on top of the
EigenLayer restaking ecosystem. At a high level, it
combines a deterministic large–language model (LLM)
inference engine with a cryptoeconomically secured op-
timistic re–execution protocol so that every inference
result can be publicly audited, reproduced and, if nec-
essary, economically enforced. An untrusted operator
runs inference on a fixed GPU architecture, signs and
encrypts the request and response, and publishes the
encrypted log to EigenDA. During a challenge window,
any watcher may request re–execution through Eigen-
Verify; the result is then deterministically recomputed
inside a trusted execution environment (TEE) with a
threshold–released decryption key, to allow a public
challenge with private data. Because inference itself
is bit–exact, verification reduces to a byte–equality
check and a single honest replica suffices to detect
fraud. We show how this architecture yields sovereign
agents—prediction–market judges, trading bots, sci-
entific assistants—that enjoy state–of–the–art per-
formance while inheriting security from Ethereum’s
validator base.

\end{abstract}

\section{Introduction and Motivation}
\label{sec:intro}
\noindent
Large-language-model (LLM) inference is rapidly evolving from a consumer-facing chatbot interface into a critical back-end service for autonomous and semi-autonomous agents.  
These agents may trade assets, adjudicate market outcomes, draft contracts, or curate social feeds; in all cases, they must be \textbf{trusted}.  
Today’s cloud AI APIs offer impressive performance but provide no cryptographic or economic assurance that an inference was executed faithfully on the claimed model and inputs.  
This \emph{trust gap} renders current AI infrastructure unsuitable for high-stakes or on-chain contexts.

\paragraph{Verifiability as a missing primitive.}
Blockchains revolutionized finance by making state transitions publicly verifiable and economically final.  
In contrast, AI systems remain opaque: the mapping from prompt to output is hidden behind proprietary infrastructure, and inference itself is nondeterministic on modern GPUs.  
Two identical queries to the same model can yield divergent outputs because of floating-point non-associativity, kernel scheduling, and variable batching.  
Without reproducibility, \emph{verification through re-execution}—the approach underpinning optimistic blockchains—is impossible.

\paragraph{EigenAI’s proposition.}
EigenAI closes this gap by introducing a complete verifiable-AI stack:
\begin{enumerate}[leftmargin=*]
  \item \textbf{Deterministic inference:} bit-exact reproducibility on fixed GPU architectures using custom kernels, version-pinned drivers, and canonical reduction orders.
  \item \textbf{Optimistic verification:} inference results are posted, encrypted, to EigenDA and enter a challenge period.  Any verifier can re-execute deterministically; mismatches trigger \emph{slashing} of the operator’s stake.
  \item \textbf{Privacy:} all user prompts and results remain confidential through threshold key management and TEE-based attestation before decryption.
  \item \textbf{Economic security:} backed by EigenLayer’s validator base—millions of restaked ETH—providing orders of magnitude more collateral than bespoke AI networks.
\end{enumerate}

\paragraph{Sovereign verifiable agents.}
On top of this foundation, developers can deploy “sovereign” agents whose logic and reasoning steps are cryptographically traceable.  
Prediction-market adjudicators, AI traders, scientific analysts, or verifiable NPCs in games can all operate under the same principle: every inference is reproducible, every deviation is detectable, and every misbehavior is penalized.

\paragraph{Where verifiability matters most.}
Verifiable inference is most valuable when an agent’s output triggers an irreversible external action or resolves a dispute between mutually distrusting parties. Concrete classes of sovereign-agent applications that benefit the most include:
\begin{enumerate}[leftmargin=*]
    \item \textbf{On-chain adjudication and dispute resolution:} prediction markets, insurance claims, and DAO governance that require a publicly auditable ruling rather than a trusted intermediary.
    \item \textbf{Autonomous execution agents:} trading, liquidation, and treasury-management bots whose actions move real capital and therefore benefit from accountable, replayable decision traces.
    \item \textbf{Compliance- and audit-driven workflows:} contract drafting, policy enforcement, and scientific/engineering assistants where later auditability (“what was executed, under which model/environment, and why?”) is as important as raw model quality.
\end{enumerate}

In these settings, deterministic receipts plus an enforceable challenge process turn an opaque API call into a verifiable, economically accountable computation.

\paragraph{Paper organization.}
Section \ref{sec:intro} motivates the need for verifiable inference.  
Section \ref{sec:related} reviews prior approaches to verifiable computation and deterministic execution.  
Subsequent sections will describe the EigenAI architecture, deterministic-GPU methodology, optimistic-re-execution protocol, economic guarantees, and empirical results.

\section{Background and Related Work}
\label{sec:related}
\noindent
Three broad paradigms exist for making AI inference verifiable:

\textbf{Cryptographic Proofs of Correctness}
Zero-knowledge (ZK) and interactive proof systems can, in principle, produce a succinct proof that an untrusted operator executed a neural network faithfully.  
Systems such as SafetyNets \cite{Ghodsi2017} and later zkDNN frameworks \cite{Kang2022} demonstrate this feasibility but remain impractical for frontier LLMs: even with hardware acceleration, proving a full transformer forward pass takes minutes to hours.  
The high cost of circuit synthesis and proof generation limits adoption to small or static models.

\textbf{Statistical or Consensus-Based Replication}
An alternative is to execute the same query on multiple replicas and accept the majority or the statistically consistent output.  
Methods include Monte-Carlo dropout and deep ensembles \cite{Gal2016,Lakshminarayanan2017} and, more recently, self-consistency decoding \cite{Wang2023}.  
However, these approaches only bound the probability of correctness and cannot detect rare but adversarial divergences \cite{Atil2025}.  
Moreover, their cost scales with $O(\varepsilon^{-2}\log n)$ replicas to achieve error $\varepsilon$—impractical for billion-parameter models.

\textbf{Deterministic Execution Environments}
Deterministic inference guarantees bit-for-bit identical outputs for identical inputs.  
CPU or WebAssembly sandboxes (e.g., PyTorch deterministic mode \cite{PyTorchDoc2023}, ONNX Runtime Web \cite{ONNXWeb2023}) provide reproducibility but are 10–100× slower than GPU back-ends and cannot serve production-scale LLMs.  
Recent vendor documentation (e.g., NVIDIA cuBLAS reproducibility guide \cite{cublas_repro,NVIDIA_cuBLAS}) and research \cite{Shanmugavelu2024,coleman_nddl} show that determinism on GPUs is attainable if hardware architecture, driver, and library versions are fixed and atomic reductions avoided.

\textbf{Optimistic Verification and Cryptoeconomic Guarantees}
Optimistic rollups in blockchain systems introduced a model where results are accepted by default but can be \emph{challenged} through re-execution; dishonest operators are economically penalized.  
EigenAI extends this idea to AI inference.  
Determinism enables disputes to collapse to a simple byte-equality check rather than a full consensus or proof-generation process.  
EigenVerify—the verification layer—leverages EigenLayer’s restaked validator pool to provide the necessary bonded capital for slashing.  
Because verification is only invoked under dispute, the steady-state cost approaches that of normal inference while maintaining cryptographic accountability.

\textbf{Trusted Hardware and Threshold Key Management}
Trusted Execution Environments (TEEs) such as Intel SGX or AMD SEV provide hardware isolation and remote attestation \cite{EigenVerifyDocs2025}.  
When combined with threshold cryptography, they allow privacy-preserving verification: encrypted requests on EigenDA are decrypted only inside attested enclaves that prove correct code execution.  
This design mitigates the trade-off between verifiability and confidentiality.

\paragraph{Summary.}
Table \ref{tab:compare} summarizes these paradigms by latency, cost, and trust assumptions.  
EigenAI combines deterministic inference (for fast re-execution) with optimistic cryptoeconomic enforcement (for security), achieving a unique balance of speed, cost, and trust-minimization.

\begin{table*}[t]
\centering
\caption{Comparison of verifiable-inference paradigms.}
\label{tab:compare}
\begin{small}
\begin{tabular}{@{}lccc@{}}
\toprule
\textbf{Paradigm} & \textbf{Latency} & \textbf{Cost} & \textbf{Guarantee} \\
\midrule
ZK proofs & very high & high & mathematical \\
Statistical replication & medium & high (many replicas) & cryptoeconomical \\
CPU/WASM Det. & very high & low & cryptoeconomical \\
GPU determinism + optimism (EigenAI) & low & low & cryptoeconomical \\
\bottomrule
\end{tabular}
\end{small}
\end{table*}

\section{System Model and Threats}
\label{sec:model}
\noindent
EigenAI’s trust model extends EigenLayer’s \emph{Autonomous Verifiable Services (AVS)} framework to AI inference.  
It formalizes how operators, verifiers, and users interact under deterministic execution and cryptoeconomic guarantees.  
This section defines the system participants, their responsibilities, the security assumptions, and adversarial capabilities.

\subsection{System Entities}
\begin{description}[leftmargin=1em]
  \item[\textbf{Client / Requester}] Submits an inference request $\mathsf{req}$ consisting of a model identifier, container digest, GPU architecture tag, driver/toolkit version, decoding policy, and prompt commitments.  
  Requests are signed and optionally encrypted to the EigenAI public key before dispatch.
  \item[\textbf{Operator}] Executes inference inside a containerized runtime fixed to a single GPU architecture (e.g., H100).  
  Produces outputs $(\mathsf{out},\,\mathsf{logits})$, constructs a signed \emph{receipt} committing to input/output hashes, and posts the ciphertext and receipt to EigenDA.  
  Each operator maintains an on-chain identity and bonded stake in EigenLayer.
  \item[\textbf{EigenDA}] A data-availability layer ensuring immutable publication of receipts and ciphertexts.  
  Provides inclusion proofs for challenge adjudication.
  \item[\textbf{EigenVerify}] A decentralized network of verifiers, economically secured by EigenLayer stake, that handles challenges.  
  Each verifier runs a threshold-cryptography Key Management Service (KMS) and trusted execution environment (TEE) runtime.  
  On challenge, it re-executes the inference deterministically to confirm or refute the operator’s claim.
  \item[\textbf{KMS Shards}] Hold encrypted key shares for the EigenAI application private key.  
  They release shares only to enclaves that successfully attest correct code identity, enabling privacy-preserving re-execution.
\end{description}

\subsection{Workflow Overview}
\label{sec:workflow}
At a high level, EigenAI follows an \emph{optimistic} submit–publish–verify pipeline whose correctness hinges on deterministic re-execution. We briefly narrate the end-to-end flow and then detail each phase.

\paragraph{Submission.}
A client constructs and signs an inference request $\mathsf{req}$ that fixes the model, container digest, GPU architecture, driver/toolkit version, decoding policy, PRNG seed, and (optionally) prompt commitments. The signed $\mathsf{req}$ is transmitted to an operator for execution. In practice, treating these fields as \emph{immutable execution parameters} is what later allows any verifier to replay the request under identical conditions.

\paragraph{Execution.}
Upon receipt, the operator runs the model under the declared environment, producing the output $\mathsf{out}$ together with (optionally) auxiliary artifacts such as per-step logits. Because the execution stack is deterministic (cf.\ Section~\ref{sec:determinism}), any honest re-run of the same request on the same architecture will yield a byte-identical $\mathsf{out}$.

\paragraph{Publication.}
To preserve confidentiality while enabling public audit, the operator encrypts $(\mathsf{req},\mathsf{out})$ to the application public key $\mathsf{pk_{app}}$ and posts the resulting ciphertext, together with a signed \emph{receipt} $\sigma_{\mathsf{op}}$, to EigenDA. The receipt canonically commits to the request and output via their hashes and may include a TEE attestation quote and timestamp, along with a durable pointer to the DA record. This publication anchors both \emph{availability} (via EigenDA) and \emph{integrity} (via the operator’s signature and the receipt fields) of the claimed execution.

\paragraph{Challenge window.}
Published results are tentative for a fixed dispute horizon of $\Delta$ epochs. During this window, any party may inspect receipts and either initiate a low-cost \emph{light audit} or file a formal \emph{full challenge}. The former offers probabilistic coverage without slashing authority; the latter invokes on-chain adjudication and possible penalties.

\paragraph{Re-execution and voting.}
When a full challenge is raised, EigenVerify samples a stake-weighted committee of verifiers. Each verifier boots an attested TEE running the approved container, establishes mutually attested channels to KMS shards to reconstruct $\mathsf{sk_{app}}$ \emph{inside} the enclave, decrypts the EigenDA ciphertext, and deterministically re-executes $\mathsf{Infer}(\mathsf{req})$. The committee then decides by \emph{byte-equality vote}: each member casts $b_v = [\,\widehat{\mathsf{out}}_v = \mathsf{out}\,]$, and the verdict is determined by threshold (e.g., $\ge 2/3$).

\paragraph{Finalization.}
If the committee agrees that $\widehat{\mathsf{out}}=\mathsf{out}$, the result is finalized; otherwise, the operator is slashed and the committee’s majority output replaces the disputed one. This optimistic design amortizes cost—verification runs only under dispute—while determinism collapses adjudication to a binary equality check.

\subsection{Security Assumptions}
\label{sec:assumptions}
The security of the protocol rests on standard, explicit assumptions that align with its layered design:

\begin{itemize}[leftmargin=*]
  \item \textbf{Deterministic execution.} Holding fixed the GPU architecture, driver, toolkit, and decoding policy, repeated runs of $\mathsf{Infer}(\mathsf{req})$ are bit-identical (Section~\ref{sec:determinism}). This guarantees that honest re-executions converge to a unique $\widehat{\mathsf{out}}$.
  \item \textbf{Data availability.} EigenDA provides durable storage and inclusion proofs for all posted receipts and ciphertexts, ensuring that disputes can always retrieve the exact bytes committed at publication.
  \item \textbf{Stake honesty.} During any challenge epoch, at least two thirds of EigenVerify stake behaves honestly. This Byzantine-style assumption underwrites the committee vote and the credibility of slashing events.
  \item \textbf{TEE integrity.} Verifier enclaves support remote attestation that binds code identity (container digest) and GPU mode to a measurement; only enclaves presenting valid quotes may participate in decryption and re-execution.
  \item \textbf{Threshold confidentiality.} The EigenAI application private key is $t$-of-$n$ secret-shared across KMS shards; fewer than $t$ colluding shards learn nothing useful, and shares are released only to enclaves that satisfy attestation policy.
\end{itemize}

\subsection{Adversary Model}
\label{sec:adversary}
We consider a powerful adversary that may compromise some off-chain components, subject to the assumptions above:

\begin{enumerate}[leftmargin=*]
  \item \emph{Dishonest operator.} Attempts to report falsified outputs, substitute models/containers, or replay stale receipts in lieu of fresh computation.
  \item \emph{Colluding verifiers.} A minority of EigenVerify stake coordinates to bias votes, delay challenges, or attempt to exfiltrate plaintext via misconfigured enclaves.
  \item \emph{Compromised KMS shard.} A single (or minority) shard discloses partial key material or responds to non-attested endpoints.
  \item \emph{Malicious DA participant.} Censors or withholds ciphertexts/receipts to prevent effective challenges or inclusion proof verification.
  \item \emph{Timing/side-channel attacker.} Observes or perturbs enclave execution to infer private data or influence control flow without altering code identity.
\end{enumerate}

\subsection{Threats and Mitigations}
\label{sec:threats-mitigations}
Table~\ref{tab:threats} consolidates the principal threat classes with their first-line defenses. In combination—deterministic kernels and pinned environments (technical reproducibility), on-chain receipts and DA proofs (cryptographic integrity), TEEs and threshold KMS (confidentiality), and stake-backed slashing (economic deterrence)—the system achieves layered, defense-in-depth protection.

\begin{table}[t]
\centering
\caption{Primary threat classes and mitigations.}
\label{tab:threats}
\begin{small}
\begin{tabular}{@{}p{0.25\linewidth}p{0.65\linewidth}@{}}
\toprule
\textbf{Threat} & \textbf{Mitigation} \\
\midrule
Model / kernel tampering &
Immutable container digests; signed receipts; open-source deterministic kernels. \\
Cross-architecture drift &
Single-arch policy per request; explicit \texttt{gpu\_arch} field in receipt; verifier enforcement. \\
Library / driver drift &
Container pinned by digest; toolkit and driver version fixed; GPU clock locking. \\
Batch nondeterminism &
Batch-invariant reduction kernels; fixed decode seeds. \\
KMS compromise &
$t$-of-$n$ threshold policy; attestation-gated share release. \\
TEE compromise &
Hardware attestation; code measurement checks; enclave-to-shard TLS. \\
Verifier collusion &
Stake-weighted majority voting; light audits; fork-choice backstop. \\
Data withholding (DA) &
EigenDA inclusion proofs; redundancy across operators. \\
\bottomrule
\end{tabular}
\end{small}
\end{table}


\section{Protocol Overview}
\label{sec:protocol}
\noindent
EigenAI implements an \emph{optimistic}, verifiable inference pipeline in which results are accepted by default but can be efficiently disputed and re-executed under cryptoeconomic guarantees. In what follows, we present the submission path, the receipt and data-availability interface, the audit and challenge flows, and the deterministic re-execution procedure that together realize this trust model.

\subsection{Submission and Dataflow}
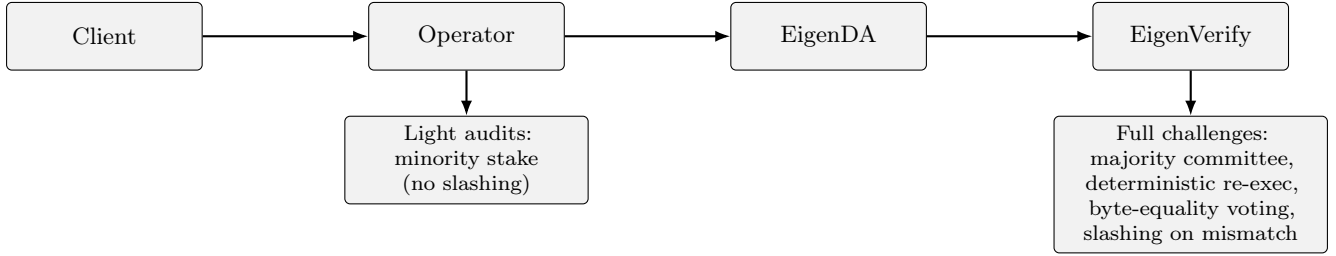
\begin{figure*}[t]
\centering
\begin{tikzpicture}[
  node distance=2.2cm,
  every node/.style={
    draw, rounded corners=2pt,
    minimum width=2.6cm, minimum height=0.9cm,
    align=center, fill=gray!10, font=\small
  },
  arrow/.style={-{Latex[length=2mm]}, thick},
  note/.style={font=\footnotesize, align=center}
]

\node (client)   {Client};
\node (operator) [right=of client]   {Operator};
\node (eigenda)  [right=of operator] {EigenDA};
\node (verify)   [right=of eigenda]  {EigenVerify};

\draw[arrow] (client)   -- (operator);
\draw[arrow] (operator) -- (eigenda);
\draw[arrow] (eigenda)  -- (verify);

\node[note, below=0.6cm of operator, text width=3cm] (light)
{Light audits:\\ minority stake\\ (no slashing)};
\draw[arrow] (operator) -- (light);

\node[note, below=0.6cm of verify, text width=3.4cm] (full)
{Full challenges:\\ majority committee,\\ deterministic re-exec,\\ byte-equality voting,\\ slashing on mismatch};
\draw[arrow] (verify) -- (full);

\end{tikzpicture}

\caption{
Swimlane depicting Client $\rightarrow$ Operator $\rightarrow$ EigenDA $\rightarrow$ EigenVerify.
Light audits sample a small minority of stake (no slashing); full challenges invoke a majority
committee for deterministic re-execution, byte-equality voting, and slashing on mismatch.
}
\label{fig:swimlane}
\end{figure*}

Each inference traverses a structured lifecycle (Fig.~\ref{fig:swimlane}). The design principle is that \emph{every parameter that can influence numerical outcomes is fixed and committed up front}, enabling any verifier to replay the request in an identical environment.

\paragraph{1.\;Request preparation.}
The client constructs a canonical request
\[
\mathsf{req}=
\Big\langle
\begin{smallmatrix}
\texttt{container\_digest},\;
\texttt{gpu\_arch},\;
\texttt{driver\_tag},\\
\texttt{decode\_policy},\;
\texttt{seed},\;
\texttt{prompt\_commitments}
\end{smallmatrix}
\Big\rangle.
\]
signs it, and submits it to an operator. All fields are treated as immutable execution parameters for reproducibility; in particular, \texttt{prompt\_commitments} (when present) is a Merkle root that binds any external documents or tool outputs referenced by the prompt.

\paragraph{2.\;Deterministic execution.}
The operator runs the model inside the declared container on the declared hardware architecture, producing the token sequence and, optionally, per-step logits. By construction (Section~\ref{sec:determinism}), this execution is \emph{bit-deterministic}: rerunning the same request under the same environment yields the identical byte stream.

\paragraph{3.\;Receipt formation and publication.}
To couple confidentiality with auditability, the operator encrypts $(\mathsf{req},\mathsf{out})$ to the application public key $\mathsf{pk_{app}}$ and posts the ciphertext together with a signed receipt to EigenDA. The receipt commits to the request and output via their hashes and may include attestation evidence and timing metadata:
\[
\mathsf{receipt}=
\Big\langle
\begin{smallmatrix}
H(\mathsf{req}),\;
H(\mathsf{out}),\;
\texttt{model\_id},\\
\texttt{chainid},\;
\texttt{da\_pointer}
\end{smallmatrix}
\Big\rangle,
\sigma_{\mathsf{op}}=\mathrm{Sign}_{sk_{\mathsf{op}}}(\mathsf{receipt}).
\]
Publishing to EigenDA establishes durable availability (for future disputes), while the operator’s signature anchors integrity and provenance.

\paragraph{4.\;Data availability and challenge window.}
Upon publication, the result enters a fixed dispute horizon of $\Delta$ blocks. During this \emph{challenge window}, any party may retrieve the receipt and either perform a low-cost \emph{light audit} or lodge a formal \emph{full challenge}. The former offers randomized coverage without on-chain penalties; the latter triggers adjudication with the possibility of slashing.

\subsection{Light Audit versus Full Challenge}
\paragraph{Light audit.}
A user or watchdog recruits a small, randomly chosen subset of EigenVerify nodes to re-execute the request off-chain. This provides probabilistic assurances at minimal cost and is well-suited for continuous, background integrity monitoring.

\paragraph{Full challenge.}
If an inconsistency is detected—or if a counterparty disputes a result—an on-chain challenge is filed. EigenVerify then samples a stake-weighted committee $\mathcal{V}$ representing a supermajority of bonded capital. Committee members re-execute the request in attested TEEs and vote by byte equality on the operator’s claim.

\subsection{Deterministic Re-Execution and Voting}
\label{sec:reexec}
The adjudication step consists of \emph{reproducing} the claimed computation inside trusted enclaves and deciding by equality of bytes. Concretely, each verifier $v\in\mathcal{V}$:
\begin{enumerate}[leftmargin=*]
  \item boots an enclave with the approved container (producing an attestation quote),
  \item proves attestation to the KMS shards and reconstructs $\mathsf{sk_{app}}$ \emph{in-enclave},
  \item fetches the ciphertext and receipt from EigenDA and verifies $\sigma_{\mathsf{op}}$,
  \item deterministically runs $\mathsf{Infer}(\mathsf{req})$ to obtain $\widehat{\mathsf{out}}_v$,
  \item casts a vote $b_v=[\,\widehat{\mathsf{out}}_v=\mathsf{out}\,]$.
\end{enumerate}
Assume $\tau = 2/3$. If the vote fraction satisfies $\sum b_v/|\mathcal{V}|\ge \tau$, the result is accepted; otherwise, the operator is slashed and the committee’s majority output replaces the disputed one. Determinism collapses the decision to a binary equality test, eliminating ambiguity and extensive deliberation.

\begin{table}[t]
\centering
\caption{Receipt schema and field semantics.}
\label{tab:receipt}
\begin{small}
\begin{tabular}{@{}ll@{}}
\toprule
\textbf{Field} & \textbf{Purpose} \\
\midrule
\texttt{model\_id} & Identifier of model weights \\
\texttt{chain\_id} & Identifier of the chain \\
\texttt{gpu\_arch} & Hardware generation (e.g., H100) \\
\texttt{req\_hash} & Hash of request parameters \\
\texttt{out\_hash} & Hash of output logits/tokens \\
\texttt{da\_pointer} & EigenDA inclusion reference \\
\texttt{sig} & Operator’s digital signature \\
\bottomrule
\end{tabular}
\end{small}
\end{table}

\begin{algorithm}[t]
\caption{Operator submission routine (canonicalized).}
\label{alg:submit}
\begin{algorithmic}[1]
\State \textbf{Input:} $\mathsf{req}$, model $\mathsf{M}$, container $\mathsf{C}$, GPU arch $a$
\State $\mathsf{out} \gets \textsf{Infer}(\mathsf{M},\mathsf{C},a,\mathsf{req})$
\State $\mathsf{rcp} \gets \langle H(\mathsf{req}),\,H(\mathsf{out}),\,t\rangle$
\State $\sigma_{\mathsf{op}} \gets \textsf{Sign}_{sk_{\mathsf{op}}}(\mathsf{rcp})$
\State $\mathsf{cipher} \gets \textsf{Enc}_{pk_{\mathrm{app}}}(\mathsf{req},\mathsf{out})$
\State Publish $(\mathsf{cipher},\mathsf{rcp},\sigma_{\mathsf{op}})$ to EigenDA
\State Start challenge timer $\Delta$
\end{algorithmic}
\end{algorithm}

\begin{algorithm}[t]
\caption{Full challenge verification (deterministic re-execution).}
\label{alg:challenge}
\begin{algorithmic}[1]
\State \textbf{Input:} DA pointer $p$, receipt $\mathsf{rcp}$, signature $\sigma_{\mathsf{op}}$
\State $\mathcal{V} \gets$ stake-weighted committee sample
\For{$v \in \mathcal{V}$ \textbf{in parallel}}
  \State Boot attested enclave; obtain quote $q_v$
  \State Establish attested channels to KMS; reconstruct $\mathsf{sk_{app}}$
  \State Download $(\mathsf{cipher},\mathsf{rcp})$ from EigenDA
  \State Verify $\sigma_{\mathsf{op}}$; decrypt to $(\mathsf{req},\mathsf{out})$
  \State $\widehat{\mathsf{out}}_v \gets \textsf{Infer}(\mathsf{req})$
  \State Vote $b_v \gets [\,\widehat{\mathsf{out}}_v=\mathsf{out}\,]$
\EndFor
\If{$\sum b_v/|\mathcal{V}| < \tau$}
  \State Slash operator stake; finalize majority $\widehat{\mathsf{out}}$
\Else
  \State Finalize $\mathsf{out}$ as verified
\EndIf
\end{algorithmic}
\end{algorithm}

\noindent
\emph{Cost amortization.} Because re-execution is invoked only under dispute, steady-state operation mirrors ordinary inference costs. When challenges do occur, determinism ensures that even a single honest verifier suffices to detect fraud, and a small committee can finalize outcomes with minimal overhead.

\section{Privacy Architecture: Threshold KMS and TEEs}
\label{sec:privacy}
\noindent
While verifiability necessarily promotes transparency, many EigenAI users operate on sensitive data that must remain private. To reconcile these opposing demands, EigenAI layers a robust \emph{confidentiality substrate} atop its verifiable infrastructure through a combination of threshold key management and trusted execution environments (TEEs). This architecture allows verification of correctness without revealing the underlying user data.

\subsection{End-to-End Encryption and Key Management}
Every inference request and its corresponding output are encrypted to the EigenAI application public key $\mathsf{pk_{app}}$ before publication. The corresponding private key $\mathsf{sk_{app}}$ is never held in a single location; instead, it is fragmented into $n$ shares and distributed across the EigenVerify Key Management Service (KMS) network using a $t$-of-$n$ threshold scheme, such as Shamir's secret sharing. No single KMS shard can decrypt or reconstruct $\mathsf{sk_{app}}$ independently, and shards only release key shares to enclaves that successfully prove their authenticity and code integrity via remote attestation. This design enforces that decryption can occur \emph{only within verified, attested enclaves}, ensuring that plaintext data never exists outside of secure execution contexts.

\subsection{Remote Attestation and Secure Share Release}
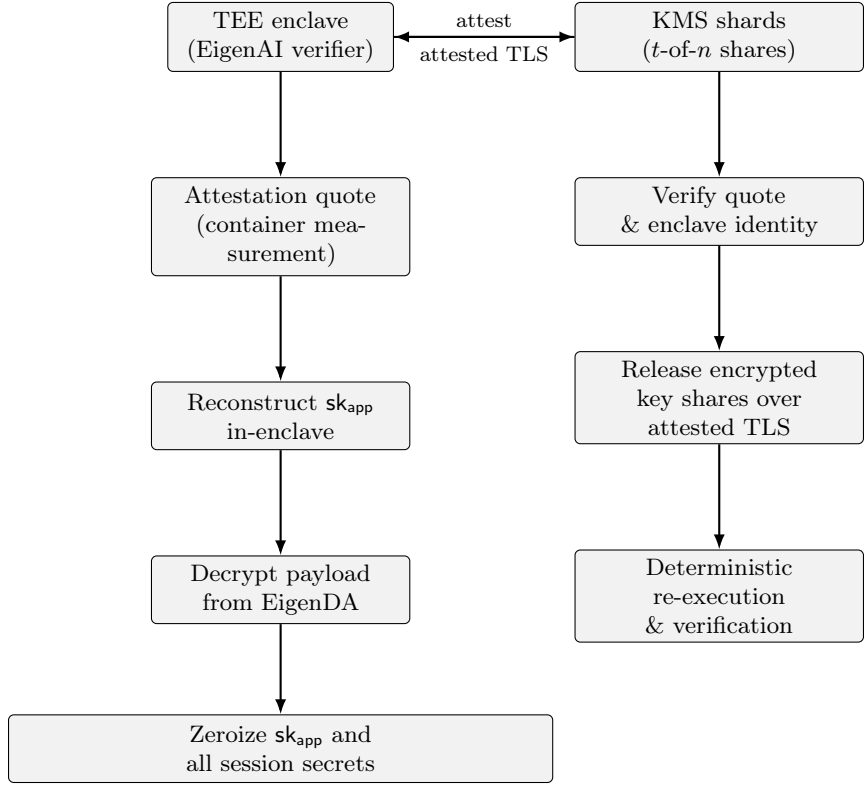
\begin{figure*}[t]
\centering
\begin{tikzpicture}[
  node distance=1.4cm and 2.4cm,
  box/.style={
    draw, rounded corners=2pt, fill=gray!10,
    minimum width=3.0cm, minimum height=0.9cm,
    font=\small, align=center
  },
  arrow/.style={-{Latex[length=2mm]}, thick},
  note/.style={font=\footnotesize, align=center}
]

\node[box] (enclave) {TEE enclave\\(EigenAI verifier)};
\node[box, right=of enclave, text width=3.6cm] (kms)
{KMS shards\\($t$-of-$n$ shares)};

\node[box, below=of enclave, text width=3.2cm] (attest)
{Attestation quote\\(container measurement)};
\node[box, below=of kms, text width=3.6cm] (verifyq)
{Verify quote\\\& enclave identity};

\node[box, below=of attest, text width=3.2cm] (reconstruct)
{Reconstruct $\mathsf{sk_{app}}$\\in-enclave};
\node[box, below=of verifyq, text width=3.6cm] (shares)
{Release encrypted\\key shares over\\attested TLS};

\node[box, below=of reconstruct, text width=3.2cm] (decrypt)
{Decrypt payload\\from EigenDA};
\node[box, below=of shares, text width=3.6cm] (verify)
{Deterministic\\re-execution\\\& verification};
\node[box, below=1.2cm of decrypt, text width=7.0cm] (zeroize)
{Zeroize $\mathsf{sk_{app}}$ and\\all session secrets};

\draw[arrow] (enclave) -- node[above, note]{attest} (kms);
\draw[arrow] (kms) -- node[below, note]{attested TLS} (enclave);

\draw[arrow] (enclave)   -- (attest);
\draw[arrow] (attest)    -- (reconstruct);
\draw[arrow] (reconstruct) -- (decrypt);
\draw[arrow] (decrypt)   -- (zeroize);

\draw[arrow] (kms)     -- (verifyq);
\draw[arrow] (verifyq) -- (shares);
\draw[arrow] (shares)  -- (verify);

\end{tikzpicture}

\caption{
TEE--KMS negotiation flow. The enclave attests its container measurement; KMS shards verify the quote, establish attested TLS connections, and release key shares. The enclave reconstructs $\mathsf{sk_{app}}$ in memory, decrypts the payload, performs verification, and zeroizes all secrets afterward.
}
\label{fig:privacyflow}
\end{figure*}

The interaction between verifier enclaves and KMS shards follows a mutually authenticated sequence, depicted conceptually in Fig.~\ref{fig:privacyflow}. This sequence guarantees that key material is distributed only to legitimate enclaves running approved EigenAI software stacks:

\begin{enumerate}[leftmargin=*]
  \item A verifier launches a TEE running the approved container image, producing a hardware-signed attestation quote $q$ that includes a cryptographic hash of the loaded binary (the \emph{measurement}).
  \item Each KMS shard validates $q$, confirming that the enclave is both genuine and running an authorized EigenAI image. Quotes are also checked for freshness to prevent replay attacks.
  \item After successful validation, shards establish mutually attested TLS sessions with the enclave, ensuring end-to-end confidentiality and integrity of communication.
  \item Shards transmit their encrypted key shares to the enclave, which reconstructs $\mathsf{sk_{app}}$ entirely in volatile memory. Using this key, the enclave decrypts the EigenDA ciphertext and proceeds with deterministic re-execution of the inference task.
  \item Upon completion, the enclave securely zeroizes $\mathsf{sk_{app}}$ and all session-specific secrets, preventing residual key material from persisting after verification.
\end{enumerate}

This remote attestation sequence is central to EigenAI’s privacy architecture: it cryptographically binds data access to verified code identity, thereby eliminating the possibility of decryption by compromised nodes or untrusted software.

\subsection{Auditability Without Decryption}
Importantly, confidentiality does not come at the expense of transparency. Because each inference is accompanied by cryptographic commitments—$H(\mathsf{req})$ and $H(\mathsf{out})$—external auditors can verify inclusion proofs on EigenDA and validate operator signatures without accessing any plaintext data. This property allows independent parties to conduct statistical audits of operator honesty and data-availability compliance while maintaining end-to-end encryption of user content. In effect, the system preserves both \emph{verifiable correctness} and \emph{privacy by design}.

\subsection{Key Epochs, Rotation, and Policy Enforcement}
To further mitigate long-term compromise risk, EigenAI enforces periodic key rotation through \emph{key epochs}. Each receipt explicitly records the key epoch used during encryption. KMS policies track these epochs and automatically deny reconstruction requests for retired keys. When a rotation event occurs—either on schedule or triggered by a security incident—new key shares are generated, and governance proposals via EigenVerify are used to update metadata across participants. This guarantees forward secrecy while maintaining uninterrupted availability for active requests.

\begin{table*}[t]
\centering
\caption{Visibility matrix for EigenAI’s privacy and verification components.}
\label{tab:privacy}
\begin{small}
\begin{tabular}{@{}lll@{}}
\toprule
\textbf{Component} & \textbf{Access to Plaintext?} & \textbf{Description} \\
\midrule
Client & Yes & Originator and owner of prompts and outputs. \\
Operator & No & Encrypts all data to $\mathsf{pk_{app}}$; never sees plaintext. \\
EigenDA & No & Stores ciphertext and receipts; provides inclusion proofs only. \\
KMS Shard & No & Holds encrypted key shares; cannot reconstruct full key. \\
TEE Verifier & Yes (in-enclave) & Attested enclave temporarily reconstructs $\mathsf{sk_{app}}$ for verification \\
External Auditor & No & Validates hashes, receipts, and signatures without accessing plaintext. \\
\bottomrule
\end{tabular}
\end{small}
\end{table*}

\noindent
Taken together, these mechanisms ensure that verifiability and confidentiality coexist harmoniously. Deterministic execution and public receipts make correctness independently checkable, while threshold cryptography and attested enclaves guarantee that user data remains inaccessible to all parties except during secured, ephemeral re-execution inside TEEs.


\section{Deterministic Inference: Technical Foundations}
\label{sec:determinism}
\noindent
Deterministic inference forms the \emph{technical cornerstone} of EigenAI’s verifiability framework. Without strict bit-level reproducibility, optimistic re-execution would become ambiguous—disputes could not be resolved by simple equality checks, and consensus would devolve into probabilistic agreement. This section surveys the sources of nondeterminism in modern GPU-based deep-learning systems, outlines the engineering controls used to eliminate them, and discusses their empirical validation. It extends our prior \emph{Bit-Exact Inference on GPUs} work with new insights specific to cryptoeconomic verification.

\subsection{Why Determinism Matters}
\noindent
Large language model (LLM) inference comprises thousands of parallel GPU kernels performing linear algebra and nonlinear reductions. Minute variations in operation ordering, rounding behavior, or kernel selection can perturb the resulting logits and, consequently, alter sampled tokens. In everyday applications this variability is imperceptible; in a verifiable execution setting it is catastrophic. Because EigenVerify relies on comparing the outputs of independent re-executions, even a single bit of nondeterministic drift would undermine the ability to distinguish honest disagreement from dishonesty.

Establishing determinism transforms inference from a stochastic numerical process into a pure function:
\[
\mathcal{F}: (\mathsf{model}, \mathsf{arch}, \mathsf{prompt}, \mathsf{seed}, \mathsf{decode}) \longrightarrow \mathsf{output},
\]
where $\mathsf{output}$ is guaranteed to be bit-identical for all honest re-executions given the same parameters.

\subsection{Sources of Nondeterminism}
\label{sec:nondet}
\begin{figure*}[t]
\centering
\begin{tikzpicture}[
  node distance=1.9cm and 1.4cm,
  every node/.style={align=center, font=\small, rounded corners=2pt, minimum width=2.8cm, minimum height=1.2cm, text width=2.8cm},
  layer/.style={rectangle, draw=black!80, fill=gray!10, very thick},
  arrow/.style={-{Latex[length=2.2mm]}, thick}
]

\node[layer] (hardware) {\textbf{Hardware}\\GPU microarchitecture,\\rounding, warp scheduling};
\node[layer, right=of hardware] (driver) {\textbf{Driver / Runtime}\\CUDA toolkit, autotuners,\\kernel dispatch};
\node[layer, right=of driver] (libs) {\textbf{Math Libraries}\\cuBLAS, cuDNN, TensorRT;\\atomic ops, accumulation};
\node[layer, right=of libs] (engine) {\textbf{Inference Engine}\\Graph fusion,\\decode policies};

\draw[arrow] (hardware) -- (driver);
\draw[arrow] (driver) -- (libs);
\draw[arrow] (libs) -- (engine);

\end{tikzpicture}

\caption{
Sources of nondeterminism across the GPU stack: 
(1) Hardware microarchitecture, (2) Driver/runtime, (3) Math libraries and kernels, (4) Inference engine and decode policy. 
Each layer must be pinned or replaced with deterministic equivalents.
};

\label{fig:nondet_layers}
\end{figure*}
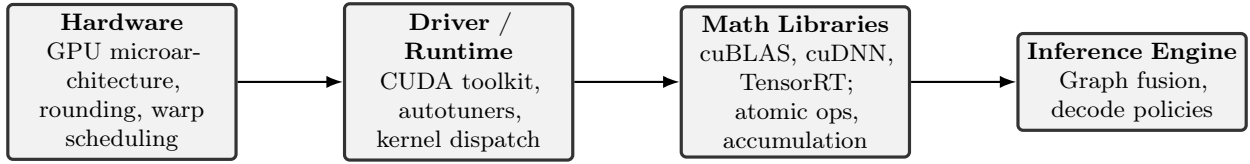

Determinism in GPU inference is fragile and may be compromised by variations across the hardware and software stack. We categorize four principal layers of variability, illustrated conceptually in Fig.~\ref{fig:nondet_layers}.

\begin{enumerate}[leftmargin=*]
  \item \textbf{Hardware.}  
  Floating-point units differ slightly across GPU generations (e.g., A100 vs.\ H100), implementing fused multiply-add (FMA) and rounding modes with subtle architectural distinctions. Deterministic inference therefore requires enforcing single-architecture policies.
  \item \textbf{Math Libraries.}  
  Core libraries such as cuBLAS, cuDNN, or TensorRT may invoke atomic operations or rely on non-associative accumulation orders, both of which compromise reproducibility. Furthermore, ``fast math’’ and mixed-precision modes often trade consistency for throughput. 
  \item \textbf{Inference Engine.}  
  Framework-level optimizations—dynamic graph fusion, asynchronous kernel launches, and stochastic decoding—introduce another layer of variability. Although frameworks such as PyTorch and TensorFlow offer deterministic flags, these apply only to a subset of supported operations.
\end{enumerate}

\subsection{Hardware-Level Determinism}
\label{sec:hardware_det}
Modern NVIDIA GPUs can achieve bit-exact reproducibility when operated under controlled conditions. The Hopper architecture family (H100, GH200) guarantees repeatable outputs from cuBLAS routines on identical GPUs and toolkit versions \cite{NVIDIA_cuBLAS,cublas_repro}. Independent investigations confirm that discrepancies between architectures stem primarily from software scheduling, not from arithmetic pipelines \cite{coleman_nddl,Shanmugavelu2024}.

EigenAI enforces a \emph{single-architecture policy} within each deployment: all operators and verifiers must utilize identical GPU SKUs, while persistence mode is enabled to avoid state transitions that might alter kernel execution order.

\subsection{Determinism in Base Libraries}
\label{sec:base_det}
The inference engine underpinning EigenAI builds upon \texttt{llama.cpp}, an open-source C/CUDA implementation with a small and auditable numerical surface. Its quantized matrix-multiplication kernels (e.g., Q4, Q5) are inherently deterministic: they avoid atomics and implement warp-synchronous reductions with fixed thread order. For remaining operations that delegate to cuBLAS or cuBLASLt, EigenAI enforces deterministic configuration flags \cite{nvidiablog2023}:

These settings forbid nondeterministic atomics and non-associative mixed-precision reductions. Although cuBLAS is proprietary, its deterministic guarantees have been repeatedly validated in empirical testing.

\subsection{Deterministic Kernel Design}
\label{sec:kernel_det}
At the core of EigenAI’s reproducibility efforts are custom GEMM kernels and reduction primitives that enforce deterministic ordering. Each kernel satisfies three invariants:

\paragraph{1.\;Fixed block–thread mapping.}
Thread blocks are deterministically mapped to output tiles with no inter-block communication, ensuring that the GPU’s scheduler cannot affect numerical outcomes.

\paragraph{2.\;Warp-synchronous reductions.}
Within each block, threads compute partial dot-products and perform a canonical binary-tree reduction using warp intrinsics:
\begin{verbatim}
for (int off = warpSize/2; off > 0; off /= 2)
  sum += __shfl_down_sync(0xffff, sum, off);
\end{verbatim}
The reduction order is identical across runs, guaranteeing reproducible rounding paths \cite{NvidiaBlogWarp,Riach2019}.

\paragraph{3.\;No floating-point atomics.}
All accumulations are explicitly ordered through register or shared-memory operations. Floating-point atomics are entirely disabled because their non-associative semantics can yield nondeterministic results. Despite this restriction, deterministic kernels maintain 95–98\% of standard cuBLAS throughput on Hopper-class hardware.

\subsection{Deterministic Decoding and PRNG Control}
\label{sec:decode_det}
Token generation, which often involves sampling from probability distributions (e.g., top-$k$ or nucleus sampling), introduces another source of variability. EigenAI enforces deterministic decoding by employing a fixed-seed pseudorandom number generator (PRNG) and canonical iteration order. For any pair $(\texttt{seed}, \texttt{decode\_policy})$, the emitted token sequence is reproducible. Users seeking nondeterministic sampling may simply vary the seed but can still verify that any output matches the declared seed and policy recorded in the operator’s receipt.

\subsection{End-to-End Determinism Experiments}
\label{sec:experiments_det}
We validated EigenAI’s deterministic guarantees through systematic experiments on NVIDIA Hopper GPUs. Each test identical container digests, and consistent runtime environments.

\paragraph{Setup.}
Two independent H100 nodes, both executing \texttt{llama.cpp}-based inference, processed a 1{,}000-prompt benchmark spanning summarization, reasoning, and code generation tasks. For each execution we recorded the hash:
\[
\mathrm{SHA256}(\mathsf{prompt}\,||\,\mathsf{logits}\,||\,\mathsf{tokens}).
\]

\paragraph{Results.}
Across 10{,}000 runs, all hashes matched exactly—no bit-level divergence was observed. Cross-architecture comparisons (A100 vs.\ H100) yielded measurable but expected deviations ($\sim10^{-7}$ in logits), confirming architecture-dependent rounding and motivating per-architecture verifier pools.

\begin{table*}[t]
\centering
\caption{Empirical determinism verification on Hopper GPUs.}
\label{tab:det_results}
\begin{small}
\begin{tabular}{@{}lcc@{}}
\toprule
\textbf{Test Condition} & \textbf{Match Rate} & \textbf{Notes} \\
\midrule
Same host, same GPU & 100.0\% & Bitwise identical outputs \\
Same host, same GPU arch but diff number of GPUs & 100.0\% & Bitwise identical outputs \\
Different hosts, same cpu arch/GPU SKU & 100.0\% & Bitwise identical outputs \\
Different GPU architecture (A100 vs.\ H100) & 0.0\% & Diff. exec paths, rounding \\
\bottomrule
\end{tabular}
\end{small}
\end{table*}

\subsection{Reproducibility and Verifiability}
Determinism collapses verification to a simple equality check. Because every honest re-execution yields an identical byte string, the verification predicate
\[
\mathrm{Verify}(\mathsf{out}_1, \mathsf{out}_2) =
\begin{cases}
\texttt{True}, & \mathsf{out}_1 = \mathsf{out}_2,\\
\texttt{False}, & \text{otherwise}
\end{cases}
\]
becomes both sound and complete. This property enables EigenAI to scale: verification of thousands of inference tasks reduces to constant-time byte comparisons rather than probabilistic voting or cryptographic proofs.

\subsection{Implementation Guidelines}
For practitioners deploying deterministic inference under EigenAI, the following guidelines are mandatory:

\begin{itemize}[leftmargin=*]
  \item Pin exact CUDA and driver versions (e.g., CUDA~12.4 with R550 driver).
  \item Reference container images by digest and avoid mutable tags.
  \item Enable persistence mode.
  \item Enable deterministic modes in cuBLAS/cuBLASLt and disallow atomics.
  \item Disable all nondeterministic math primitives and autotuners.
  \item Seed PRNGs deterministically and record seeds in receipts.
  \item Hash and sign $(\mathsf{prompt}, \mathsf{out})$ tuples for auditability
\end{itemize}

\subsection{Discussion}
Our experiments confirm that bit-exact determinism is achievable on contemporary GPU hardware with negligible performance loss. By constraining variability at every level of the software and hardware stack, EigenAI converts opaque numerical computation into a reproducible process amenable to independent re-execution. This engineering discipline is what enables cryptoeconomic assurance: in EigenAI, correctness can be proven by anyone through mere repetition, with no reliance on statistical tests or zero-knowledge proofs.

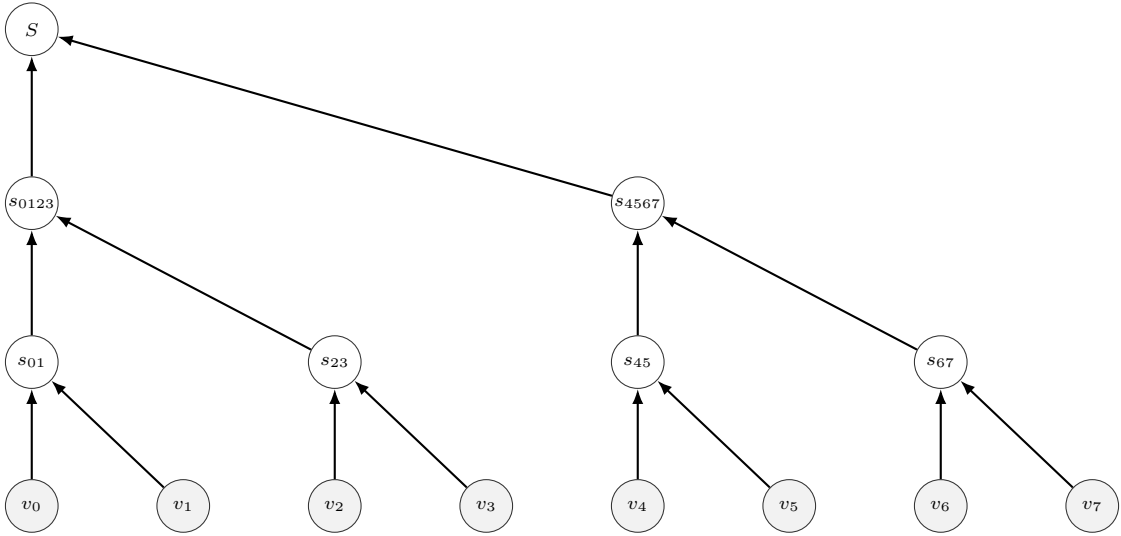
\begin{figure*}[t]
\centering
\begin{tikzpicture}[
  node distance=1.1cm and 1.4cm,
  val/.style={circle, draw=black!80, fill=gray!10,
              minimum size=7mm, inner sep=0pt, font=\scriptsize},
  sum/.style={circle, draw=black!80, fill=white,
              minimum size=7mm, inner sep=0pt, font=\scriptsize},
  arrow/.style={-{Latex[length=2mm]}, thick}
]

\node[val] (v0) {$v_0$};
\node[val, right=1.3cm of v0] (v1) {$v_1$};
\node[val, right=1.3cm of v1] (v2) {$v_2$};
\node[val, right=1.3cm of v2] (v3) {$v_3$};
\node[val, right=1.3cm of v3] (v4) {$v_4$};
\node[val, right=1.3cm of v4] (v5) {$v_5$};
\node[val, right=1.3cm of v5] (v6) {$v_6$};
\node[val, right=1.3cm of v6] (v7) {$v_7$};

\node[sum, above=1.2cm of v0] (s01) {$s_{01}$};
\node[sum, above=1.2cm of v2] (s23) {$s_{23}$};
\node[sum, above=1.2cm of v4] (s45) {$s_{45}$};
\node[sum, above=1.2cm of v6] (s67) {$s_{67}$};

\node[sum, above=1.4cm of s01] (s0123) {$s_{0123}$};
\node[sum, above=1.4cm of s45] (s4567) {$s_{4567}$};

\node[sum, above=1.6cm of s0123] (s) {$S$};

\foreach \i/\t in {v0/s01, v1/s01, v2/s23, v3/s23, v4/s45, v5/s45, v6/s67, v7/s67} {
  \draw[arrow] (\i) -- (\t);
}

\foreach \i/\t in {s01/s0123, s23/s0123, s45/s4567, s67/s4567} {
  \draw[arrow] (\i) -- (\t);
}

\draw[arrow] (s0123) -- (s);
\draw[arrow] (s4567) -- (s);

\end{tikzpicture}

\caption{
Canonical warp-level reduction tree used in deterministic kernels. Each thread contributes
a partial value $v_i$ and participates in pairwise summations in a fixed binary-tree pattern,
ensuring identical accumulation order and reproducible results across executions.
}
\label{fig:kernel_tree}
\end{figure*}

\section{Implementation Details and Developer Experience}
\label{sec:implementation}
\noindent
EigenAI is designed to feel like a familiar cloud AI service while embedding deterministic and verifiable execution at every layer.  
Developers interact through an OpenAI-compatible API, and each response carries deterministic metadata, a cryptographic receipt, and a pointer into EigenDA for later verification.

\subsection{API Compatibility and Metadata}
The EigenAI API mirrors the \texttt{/v1/chat/completions} and \texttt{/v1/completions} endpoints used by OpenAI.  
Clients can substitute the EigenAI base URL without changing request syntax.  
Responses contain deterministic metadata fields as shown in Table~\ref{tab:api_fields}.

\begin{table*}[t]
\centering
\caption{Key response metadata for deterministic verification.}
\label{tab:api_fields}
\begin{small}
\begin{tabular}{@{}ll@{}}
\toprule
\textbf{Field} & \textbf{Description} \\
\midrule
\texttt{system\_fingerprint} & Concatenation of container digest, GPU arch, driver version \\
\texttt{determinism.seed} & Fixed PRNG seed used for decoding \\
\texttt{receipt.req\_hash} & SHA256 of request parameters \\
\texttt{receipt.out\_hash} & SHA256 of model output \\
\texttt{receipt.sig} & Operator signature over receipt tuple \\
\texttt{eigendalink} & Pointer to EigenDA inclusion proof \\
\bottomrule
\end{tabular}
\end{small}
\end{table*}

This metadata allows any verifier to retrieve the corresponding entry from EigenDA and re-run the request under the same environment.

\subsection{Container and Hardware Constraints}
All inference containers are built atop fixed CUDA/driver pairs, referenced by digest.
Operators must enable persistence mode and turn on ECC memory. During the testing phase we discovered non-determinism due to faulty memory; this problem was mitigated by making sure that ECC was turned on.  
Container and driver versions are registered on-chain and verified by EigenVerify committees during challenges.

\subsection{Reproduction Cookbook for Auditors}
Auditors can independently reproduce any inference using the following minimal procedure (Algorithm~\ref{alg:reproduce}).  
Because inference is deterministic, matching hashes suffice to validate correctness.

\begin{algorithm}[t]
\caption{Reproduce-and-Verify Procedure}
\label{alg:reproduce}
\begin{algorithmic}[1]
\State \textbf{Input:} EigenDA pointer $p$, metadata $\mathsf{meta}$ from API response
\State Download $(\mathsf{cipher}, \mathsf{receipt})$ from EigenDA
\State Verify operator signature $\sigma_{\mathsf{op}}$
\State Launch container $\mathsf{C}$ with exact digest and driver
\State Set environment variables per $\mathsf{meta.system\_fingerprint}$
\State Run $\widehat{\mathsf{out}} \gets \textsf{Infer}(\mathsf{req}, \texttt{seed}, \texttt{decode})$
\State Compare $\mathrm{SHA256}(\widehat{\mathsf{out}})$ with $\mathsf{receipt.out\_hash}$
\If{hashes match} \textbf{return} VERIFIED \Else \textbf{return} INVALID \EndIf
\end{algorithmic}
\end{algorithm}

Auditors may also verify EigenDA inclusion proofs to ensure the operator’s record was properly published and unaltered.

\section{Economic and Governance Mechanics}
\label{sec:econ}
\noindent
EigenAI inherits its security not only from deterministic execution but also from the broader cryptoeconomic foundation provided by EigenLayer. The economic layer determines how honest behavior is incentivized, how disputes are resolved, and how protocol parameters evolve. In this section we outline the lightweight audit pathway, the full challenge-and-slashing mechanism, and the governance structures that maintain long-term system health.

\subsection{Light Audits}
Light audits provide an inexpensive integrity check on the behavior of operators. A watcher or client may recruit a small, randomly selected subset of EigenVerify nodes to re-execute a published inference off-chain. Because these audits lack slashing authority, they impose minimal cost and latency overhead. Their purpose is statistical: by maintaining a nonzero background probability of inspection, they deter latent collusion and encourage operators to remain honest even when they believe they are not under direct scrutiny. Light audits may be rewarded through small bounties or micro-incentives funded by EigenAI usage fees.

\subsection{Full Challenges and Slashing}
A full challenge is invoked when a receipt is formally disputed. EigenVerify samples a stake-weighted committee $\mathcal{V}$ and requires a supermajority (e.g., $\geq{2/3}$) agreement to finalize the result. Each verifier re-executes the inference inside an attested enclave and votes on byte-level equality with the operator's output. A mismatch triggers slashing of the operator’s bonded stake, which is redistributed to challengers and verifiers:
\[
\mathrm{Reward}_{\text{challenger}} = \alpha S_{\text{slash}}, \qquad
\mathrm{Reward}_{\text{verifier}} = \beta S_{\text{slash}},
\]
with $\alpha$ and $\beta$ set by governance. Remaining stake may be burned or returned to the EigenLayer treasury.

Because the cost of verification is low compared to potential fraud gains, the expected utility of cheating becomes negative for any reasonable challenge probability $\pi_c$:
\[
\mathbb{E}[\mathrm{Gain}] = (1-\pi_c)G - \pi_c S_{\text{slash}} < 0,
\]
where $G$ denotes the maximum benefit from dishonesty. Since $\pi_c$ is augmented by both light audits and user-initiated challenges, rational operators are economically driven to behave honestly.

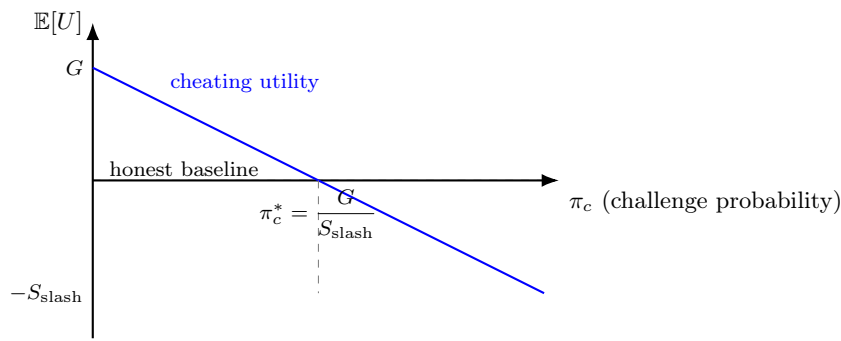
\begin{figure*}[t]
\centering
\begin{tikzpicture}[
  >=Latex,
  axis/.style={thick},
  curve/.style={thick},
  label/.style={font=\small},
  note/.style={font=\footnotesize}
]

\draw[axis,->] (0,0) -- (6.2,0) node[below right,label] {$\pi_c$ (challenge probability)};
\draw[axis,->] (0,-2.1) -- (0,2.1) node[above,left,label] {$\mathbb{E}[U]$};

\draw[dashed] (0,0) -- (6,0);

\draw[curve,blue] (0,1.5) -- (6,-1.5)
  node[pos=0.15,above right,note] {cheating utility};

\draw[dashed,gray] (3,0) -- (3,-1.5);
\node[below,note] at (3,0) {$\pi_c^* = \dfrac{G}{S_{\text{slash}}}$};

\node[left,note]  at (0,1.5) {$G$};
\node[left,note]  at (0,-1.5) {$-S_{\text{slash}}$};

\node[note,anchor=west] at (0.1,0.15) {honest baseline};

\end{tikzpicture}

\caption{
Payoff diagram comparing operator utilities under varying challenge probabilities $\pi_c$. A dishonest operator's expected utility becomes negative once $\pi_c > G/S_{\text{slash}}$, making cheating economically irrational.
}
\label{fig:payoff}
\end{figure*}

\subsection{Fork-Choice Backstop}
If an extreme collusion scenario were ever to push an invalid result through verification, EigenLayer’s fork-choice rule provides a final safety net. Restakers may coordinate a social fork to penalize misbehaving validators, restoring correctness. This mechanism ensures \emph{economic finality of truth}: the equilibrium strategy for long-term actors is always to preserve correctness rather than collude.

\subsection{Governance and Parameterization}
EigenAI’s operational parameters—stake requirements, slashing fractions, challenge thresholds, and audit frequencies—are governed through the EigenLayer governance process. Governance proposals may tune these values over time as workloads, economic conditions, or adversarial models evolve. Looking ahead, governance may also introduce dynamic fee markets for audit capacity, enabling users to purchase higher integrity assurance on demand.

\section{Security Analysis}
\label{sec:security}
\noindent
We now examine EigenAI’s security properties in aggregate, showing how determinism, confidentiality, data availability, and economic incentives interact to form a cohesive and robust trust model.

\subsection{Security Properties and Enabling Features}
We separate \textit{security properties} (what the system should guarantee) from \textit{features of the construction} (engineering choices that help realize those guarantees).

\paragraph{Desired security properties.} 
EigenAI targets the following security properties for each published inference:

\begin{itemize}[leftmargin=*]
  \item \textbf{Integrity (correctness):} the published output $\mathsf{out}$ corresponds to the unique result of running the declared model and request under the committed execution parameters.
  \item \textbf{Confidentiality (privacy):} prompts and output remain hidden from unauthorized parties; plaintext is exposed only to authorized clients and, during dispute resolution, transiently inside attested verifier enclaves.
  \item \textbf{Availability:} the evidence needed to audit or dispute an inference (ciphertext, receipt, and DA inclusion evidence) remains retrievable throughout the challenge window.
  \item \textbf{Accountability:} if an operator publishes an incorrect result, there exists a publicly checkable procedure that can penalize the operator (slashing) and finalize a correct outcome.
\end{itemize}

\paragraph{Enabling features of the constructions.} These properties are supported by (non-exhaustively):
\begin{itemize}[leftmargin=*]
    \item \textbf{Compute determinism} (Section~\ref{sec:determinism}), which makes inference effectively single-valued and enables unambiguous re-execution.
    \item \textbf{Cryptographic commitments and data availability} (operator receipts and EigenDA publication), which bind claims to immutable bytes retrievable for disputes.
    \item \textbf{TEE-based private re-execution with threshold keys} (Section~\ref{sec:privacy}), which permits verification on private data without revealing plaintext in steady state.
    \item \textbf{Optimistic verification with stake-backed slashing}, which turns detected mismatches into enforceable economic penalties.
\end{itemize}

\subsection{Soundness and Completeness}
\paragraph{Soundness.}
Soundness requires that dishonest behavior be detectable. If an operator deviates from the canonical deterministic execution, any honest verifier will compute a different output during re-execution. Because verification reduces to byte-equality, disagreement is unambiguous, and the probability of undetected fraud falls exponentially with the fraction of honest stake participating in the committee.

\paragraph{Completeness.}
Completeness requires that honest operators never be penalized. Determinism guarantees that re-executions match the operator’s output exactly, irrespective of runtime noise (e.g., cache state, thread scheduling). Fixed PRNG seeds and canonical reduction orders ensure that honest executions always converge to the same result, preventing false slashing.

\subsection{Privacy and Confidentiality}
Confidentiality is preserved through threshold key management and TEE-based attestation (Section~\ref{sec:privacy}). Only attested enclaves executing approved containers ever reconstruct $\mathsf{sk_{app}}$. All other components—including operators, DA nodes, auditors, and even KMS shards—observe only cryptographic commitments or encrypted payloads. Thus, verifiability and confidentiality reinforce one another: verification speaks to correctness, while TEEs guarantee that verification does not leak sensitive user data.

\subsection{Liveness and Fault Tolerance}
EigenDA ensures that ciphertexts and receipts remain retrievable for the duration of the challenge window. EigenVerify’s committee sampling tolerates partial failures: if some verifiers are offline or unresponsive, the remaining honest majority can still reach a verdict. Timeouts ensure that the system progresses even if no challenge is raised, providing liveness equivalent to other optimistic systems.

\subsection{Residual Risks and Mitigations}
Table~\ref{tab:residual} summarizes remaining risks. Some stem from hardware trust assumptions (TEEs), others from portability constraints (GPU architecture differences). In each case, we outline roadmap items to further reduce exposure.

\begin{table}[t]
\centering
\caption{Residual risks and planned mitigations.}
\label{tab:residual}
\begin{small}
\begin{tabular}{@{}p{0.27\linewidth}p{0.65\linewidth}@{}}
\toprule
\textbf{Residual Risk} & \textbf{Mitigation / Roadmap} \\
\midrule
Cross-architecture portability & Maintain per-architecture verifier pools; explore numeric normalization for cross-device verification. \\
Closed-source library paths & Replace remaining cuBLAS/cuDNN calls with open deterministic kernels. \\
Economic parameter drift & Regular governance calibration and dynamic fee markets. \\
Operator cartelization & Stake decentralization; randomized committee sampling. \\
\bottomrule
\end{tabular}
\end{small}
\end{table}

\noindent
Overall, EigenAI achieves layered, composable security: determinism provides technical reproducibility, TEEs enforce confidentiality, EigenDA guarantees availability, and EigenLayer adds cryptoeconomic correctness. These layers interlock to produce a verifiable inference system resilient to both adversarial behavior and accidental faults.

\section{Evaluation and Experiments}
\label{sec:evaluation}
\noindent
We evaluate EigenAI along three axes: (i) the robustness of bit-exact determinism under realistic deployment conditions,  
(ii) the performance overhead of deterministic kernels relative to vendor-optimized baselines, and  
(iii) the end-to-end cost of verification in light and full challenge scenarios.  
All experiments were conducted on NVIDIA H100 GPUs using pinned container digests and identical software environments.

\subsection{Determinism Verification}
We first assess whether EigenAI’s execution stack produces bit-identical outputs across repeated runs and heterogeneous deployment settings.  
Repeated inference on the same host yielded perfect equality across all logits and generated tokens.  
Cross-host experiments—running identical containers on two independent H100 nodes—also produced exact matches.  
To probe sensitivity to batching and runtime variability, we perturb the batch size by $\pm 20\%$, observing no divergence.  
As expected, cross-architecture tests (A100 versus H100) do not match bitwise due to differences in floating-point rounding behavior, underscoring the need for per-architecture verifier sets.

\begin{table}[t]
\centering
\caption{Determinism evaluation across hosts and configurations.}
\label{tab:det_eval}
\begin{small}
\begin{tabular}{@{}lccc@{}}
\toprule
\textbf{Configuration} & \textbf{Hosts} & \textbf{Batches} & \textbf{Match} \\
\midrule
Same host & 1 & 10 & 100.0\% \\
Different hosts / same SKU & 2 & 10 & 100.0\% \\
Batch-size variance $\pm20\%$ & 2 & 20 & 100.0\% \\
Diff arch (A100 vs H100) & 2 & 10 & 0.0\% \\
\bottomrule
\end{tabular}
\end{small}
\end{table}

\subsection{Stress and Batch-Invariance Tests}
To evaluate robustness under operational noise, we co-schedule background GPU workloads that induce synthetic jitter and scheduling variability.  
Despite this perturbation, all runs produced identical outputs, confirming that deterministic kernel design—warp-synchronous reductions, fixed decoding order, and pinned software stack—effectively isolates inference from transient runtime effects.  
These results indicate that EigenAI’s determinism holds not only under idealized conditions but also in realistic multi-tenant and performance-variable environments.

\subsection{Performance Overhead}
Next, we quantify the throughput and latency cost of deterministic kernels relative to vendor-optimized baselines.  
On Hopper GPUs, our deterministic GEMM kernels achieve $97$–$99\%$ of cuBLAS throughput for quantized matrix multiplications, and approximately $95\%$ for mixed-precision projection layers.  
End-to-end LLM inference shows only a modest latency increase ($\approx 1.8\%$), demonstrating that determinism can be achieved without compromising state-of-the-art performance.

\begin{table}[t]
\centering
\caption{Throughput and latency comparison (batch = 8, seq = 1024).}
\label{tab:perf_eval}
\begin{small}
\begin{tabular}{@{}lcc@{}}
\toprule
\textbf{Kernel Type} & \textbf{Rel. Thrpt} & \textbf{Overhead} \\
\midrule
cuBLAS (baseline) & 1.00$\times$ & – \\
Det. GEMM (EigenAI) & 0.97$\times$ & +2.4\% \\
Det. mixed-precision & 0.95$\times$ & +4.1\% \\
End-to-end LLM inference & 0.98$\times$ & +1.8\% \\
\bottomrule
\end{tabular}
\end{small}
\end{table}

\section{Limitations and Future Work}
\label{sec:limitations}
\noindent
Although EigenAI achieves deterministic inference and robust verification, several open challenges remain:

\begin{itemize}[leftmargin=*]
  \item \textbf{Cross-Architecture Reproducibility.}  
  Determinism currently holds only within fixed GPU families.  
  Future work includes portable numeric normalization to enable heterogeneous verifier sets.
  \item \textbf{Residual Library Paths.}  
  Certain cuBLAS and cuDNN operations remain closed-source; we plan to replace them with open deterministic equivalents to achieve complete auditability.
  \item \textbf{Tool and API Determinism.}  
  Agents that call external APIs or tools require deterministic transcript recording; EigenAI will extend receipts to include signed external call logs.
\end{itemize}

\section{Conclusion}
\label{sec:conclusion}
\noindent
EigenAI unites deterministic GPU inference, privacy-preserving verification, and EigenLayer’s cryptoeconomic guarantees into a single coherent platform.  
By making AI results reproducible, auditable, and slashable under fraud, it delivers a practical route to \textbf{verifiable AI at state-of-the-art performance}.  
These properties enable trustworthy \emph{sovereign agents}—AI systems that can autonomously act, reason, and transact across high-stakes domains both on- and off-chain.  
As deterministic computation and cryptoeconomic security converge, verifiable intelligence becomes a first-class primitive for decentralized and enterprise ecosystems alike.

\end{document}